\documentclass[prd,twocolumn,superscriptaddress,nofootinbib]{revtex4-2}

\usepackage{graphicx}
\usepackage{epsfig}
\usepackage{epsf,amsfonts}
\usepackage{float}

\usepackage{amsmath}
\usepackage{amssymb}
\usepackage{amsfonts}
\usepackage{latexsym}
\usepackage{amsthm}
\usepackage{mathtools}
\usepackage{bbm}
\usepackage{mathrsfs}

\usepackage[T1]{fontenc}
\usepackage{slashed}
\usepackage{esvect}

\usepackage[usenames,dvipsnames]{xcolor}
\usepackage[colorlinks=true,citecolor=Blue,linkcolor=RubineRed,urlcolor=Blue]{hyperref}

\usepackage{dcolumn}
\usepackage{bm}

\begin{document}

\title{Axial-anomaly effects and chiral phase structure in holographic QCD}

\author{Xin-Yi Liu}
\email{liuxinyi23@mails.ucas.ac.cn}
\affiliation{School of Fundamental Physics and Mathematical Sciences, Hangzhou Institute for Advanced Study, UCAS, Hangzhou 310024, China}
\affiliation{Institute of Theoretical Physics, Chinese Academy of Sciences, Beijing 100190, China}
\affiliation{University of Chinese Academy of Sciences (UCAS), Beijing 100049, China}
\author{Yue-Liang Wu}
\email{ylwu@itp.ac.cn}
\affiliation{School of Fundamental Physics and Mathematical Sciences, Hangzhou Institute for Advanced Study, UCAS, Hangzhou 310024, China}
\affiliation{Institute of Theoretical Physics, Chinese Academy of Sciences, Beijing 100190, China}
\affiliation{University of Chinese Academy of Sciences (UCAS), Beijing 100049, China}
\affiliation{International Center for Theoretical Physics Asia-Pacific (ICTP-AP), UCAS, Beijing 100190, China}
\author{Zhen Fang}
\email{zhenfang@hnu.edu.cn}
\affiliation{School for Theoretical Physics, School of Physics and Electronics, Hunan University, Changsha 410082, China}
\affiliation{Hunan Provincial Key Laboratory of High-Energy Scale Physics and Applications, Hunan University, Changsha 410082, China}

\begin{abstract}

We study the impact of axial-anomaly effects on the chiral phase structure in a \(U(3)\)-extended soft-wall holographic QCD model. Including the pseudoscalar singlet sector allows for a dynamical description of the \(\eta\)-\(\eta^\prime\) system through a determinant interaction with a holographic-coordinate-dependent strength. Vacuum pseudoscalar observables, particularly the \(\eta^\prime\) mass and the \(\eta\)-\(\eta^\prime\) mixing pattern, constrain the overall magnitude of the anomaly contribution but leave its holographic profile largely undetermined. We then examine how different anomaly profiles consistent with vacuum phenomenology affect the finite-temperature chiral transition. Constructing the Columbia plot within this framework, we find that the predicted phase structure depends sensitively on the anomaly implementation: some profiles yield crossover/second-order behavior across the entire quark-mass plane, while others generate a first-order region in the light-quark corner. These results highlight the strong sensitivity of the holographic QCD phase structure to the modeling of axial-anomaly effects. 

\end{abstract}

\maketitle

\section{Introduction}

Understanding the chiral phase structure of quantum chromodynamics (QCD) remains one of the central challenges in strong-interaction physics \cite{Stephanov:2004wx,Busza:2018rrf}. The restoration of chiral symmetry at high temperature plays a key role in the dynamics of strongly interacting matter created in relativistic heavy-ion collisions and in the evolution of the early Universe. In particular, the nature of the chiral phase transition and its dependence on the light and strange quark masses, commonly summarized in the Columbia plot \cite{Brown:1990ev}, has been the subject of extensive analytical, numerical, and phenomenological investigations \cite{Ding:2015ona,Resch:2017vjs,Fischer:2018sdj,Aarts:2015tyj}. Despite the substantial progress achieved in lattice QCD, especially at vanishing baryon chemical potential, important questions persist concerning the detailed structure of the phase diagram and its sensitivity to different symmetry-breaking mechanisms.

An important aspect of this problem concerns the role of axial $U(1)_A$ symmetry breaking in shaping the chiral phase structure \cite{Pisarski:1983ms,deForcrand:2017cgb,Cuteri:2021ikv,Dini:2021hug,Fejos:2022mso,Pisarski:2024esv,Giacosa:2024orp,Fejos:2024bgl}. Although the axial symmetry is explicitly broken in QCD by the quantum anomaly, its manifestation in effective descriptions typically relies on phenomenological interactions. The precise form of such interactions is not uniquely determined and may vary significantly among different model frameworks. 
Consequently, it remains an important theoretical question to clarify how different realizations of $U(1)_A$-breaking effects may influence the predicted chiral phase structure and the resulting Columbia plot.

Holographic QCD models provide a complementary framework for exploring the impact of phenomenological symmetry-breaking mechanisms on hadron properties and chiral dynamics in strongly coupled regimes. Bottom-up constructions inspired by the AdS/CFT correspondence have been widely employed to investigate chiral symmetry breaking, hadron spectra, and finite-temperature phase transitions \cite{DaRold:2005mxj,Erlich:2005qh,Karch:2006pv,Herzog:2006ra,Gursoy:2007cb,Gursoy:2007er,Cherman:2008eh,Gubser:2008yx,Fujita:2009wc,Colangelo:2009ra,DeWolfe:2010he,Jarvinen:2011qe,Alho:2012mh,Rougemont:2015ona,Li:2022erd,Rougemont:2023gfz,Liu:2023pbt,Cai:2024eqa,CruzRojas:2024etx,Liu:2024efy,Jokela:2024xgz,Ecker:2025vnb,Tong:2025rxz,Bartolini:2025sag,Chen:2025ncu,Shen:2025zkj,Shen:2025yrn,Hippelainen:2026izm,Deng:2026aht}. Among these approaches, soft-wall models \cite{Karch:2006pv} offer a particularly economical and flexible framework in which confinement-related features and linear Regge trajectories can be incorporated while maintaining both analytical transparency and numerical tractability. Various extensions of the soft-wall framework have achieved notable success in describing hadron spectra and aspects of chiral thermodynamics \cite{Katz:2007tf,Sui:2009xe,Sui:2010ay,Li:2013oda,Cui:2013xva,Cui:2014oba,Chelabi:2015cwn,Li:2016smq,Fang:2016uer,Fang:2016dqm,Fang:2016cnt,Critelli:2017oub,Fang:2019lsz,Giannuzzi:2021euy,Cao:2021tcr,Cao:2022csq,Colangelo:2023een,Chen:2024cxh,Zheng:2024rzl,Shen:2026bko}.

In most existing soft-wall holographic studies of the chiral phase structure, however, the flavor sector is restricted to $SU(2)$ or $SU(3)$, and anomaly effects associated with the axial $U(1)_A$ symmetry are either absent or implemented only in a simplified manner. As a consequence, the pseudoscalar singlet sector, in particular the $\eta'$ meson, is typically absent or not dynamically described. This limitation restricts the ability of such models to investigate how different phenomenological realizations of $U(1)_A$-breaking effects, constrained by vacuum pseudoscalar observables, may influence finite-temperature chiral dynamics and the Columbia plot.

In this work, we construct a \(U(3)\) extension of our previous soft-wall holographic QCD model \cite{Fang:2019lmd}. The purpose of this extension is to investigate the impact of a different and explicit \(U(1)_A\)-breaking interaction, controlled by a phenomenological function \(\gamma(z)\). This term enables a dynamical treatment of the pseudoscalar singlet channel and induces nontrivial mixing between the flavor-singlet and octet components. As a result, the \(\eta\)-\(\eta^\prime\) mass splitting and the corresponding mixing structure emerge naturally from the coupled equations of motion, rather than being imposed by hand.

We first use the vacuum pseudoscalar spectrum, in particular the \(\eta^\prime\) mass and the \(\eta\)-\(\eta^\prime\) mixing structure, to constrain the allowed range of the anomaly-related parameter \(\gamma(z)\). It is important to emphasize that \(\gamma(z)\) should be interpreted as an effective function encoding explicit \(U(1)_A\) breaking within the present bottom-up framework, rather than as a direct measure of microscopic topological observables. In particular, no attempt is made here to compute quantities such as the topological susceptibility or axial susceptibilities. Instead, the focus of the present work is on the phenomenological consequences of varying the effective anomaly strength within a unified holographic model.

Having determined the model parameters from vacuum spectroscopy, we subsequently investigate how the same parameter influences the finite-temperature chiral phase structure. In particular, we analyze the dependence of the Columbia plot on \(\gamma(z)\), examining how the order of the chiral transition and the boundaries separating first-order and crossover regions respond to variations in the effective \(U(1)_A\)-breaking strength. This allows us to explore, within a controlled holographic framework, how vacuum \(\eta^\prime\) physics and thermal chiral dynamics are interconnected.

The central objective of this work is therefore not to determine the microscopic mechanism responsible for axial symmetry breaking in QCD. Rather, our goal is to investigate how different effective anomaly profiles, constrained by vacuum pseudoscalar observables such as the \(\eta^\prime\) mass and the \(\eta\)-\(\eta^\prime\) mixing structure, influence the predicted chiral phase structure within a holographic framework. This provides a controlled setting in which to examine the sensitivity of the Columbia plot to phenomenological implementations of anomaly effects and highlights the importance of consistently connecting vacuum hadron physics with finite-temperature dynamics in effective QCD models.

The remainder of this paper is organized as follows. In Sec.~\ref{Sec: Introduce the model}, we introduce the \(U(3)\) soft-wall holographic model and specify the form of the \(U(1)_A\)-breaking interaction. Sec.~\ref{Sec: Pseudoscalar sector} is devoted to the analysis of the vacuum pseudoscalar sector, including \(\eta\)-\(\eta^\prime\) mixing and the resulting constraints on \(\gamma(z)\). In Sec.~\ref{Sec: Scalar sector}, we extend the discussion to the scalar sector and examine the additional constraints that arise from scalar meson observables. Sec.~\ref{Sec: Finite T} presents our analysis of the finite-temperature chiral phase structure, including the corresponding Columbia plots for different choices of \(\gamma(z)\). Finally, we summarize our main results and provide an outlook in Sec.~\ref{Sec: Conclusion}.

\section{Model setup}\label{Sec: Introduce the model}

\subsection{Soft-wall holographic framework}

Our analysis is based on the soft-wall holographic QCD model developed in Ref.~\cite{Fang:2019lmd}. 
That framework successfully captures spontaneous chiral symmetry breaking and reproduces the octet meson spectra within an \(SU(3)_L\times SU(3)_R\) flavor structure, but it does not incorporate the \(\eta^\prime\) meson. Since the present work represents a direct \(U(3)\) extension of that model, we only briefly summarize the essential ingredients needed to establish our notation, referring the reader to Ref.~\cite{Fang:2019lmd} for a detailed exposition.

The five-dimensional background geometry is taken to be AdS\(_5\),
\begin{align}
ds^2 = e^{2A(z)}\left(\eta_{\mu\nu}dx^\mu dx^\nu - dz^2\right),
\end{align}
with the warp factor given by \(A(z) = -\log\frac{z}{L}\). The AdS radius will be set to \(L=1\) throughout this work. Chiral dynamics is introduced through a bifundamental scalar field \(X\), together with \(U(3)_L \times U(3)_R\) gauge fields \(A_{L,R}\) that encode the chiral flavor symmetry of the boundary theory.

The bulk action takes the form
\begin{align}\label{Eq: action}
S= &\int d^5 x \sqrt{g}\, e^{-\Phi(z)}
\left[
\operatorname{Tr}\left\{
|D X|^2
-m_5^2(z)|X|^2
-\lambda |X|^4
\right.\right.\notag\\
&\quad \left.\left.
-\frac{1}{4 g_5^2}
\left(F_L^2+F_R^2\right)
\right\}
-\gamma(z) \operatorname{Re}\{\det X\}
\right],
\end{align}
where the covariant derivative is defined as \(D^MX = \partial^M X - i A_L^M X + i X A_R^M\). 
The five-dimensional gauge coupling is fixed by matching to QCD as
\begin{align}
g_5^2 = \frac{12\pi^2}{N_c},
\end{align}
with \(N_c\) denoting the number of colors \cite{Erlich:2005qh}. 

The dilaton profile and the scalar bulk mass are chosen as
\begin{align}
\Phi(z)=\mu_g^2 z^2\left(1-e^{-\frac{1}{4}\mu_g^2 z^2}\right),
\quad
m_5^2(z)=-3-\mu_c^2 z^2 ,
\end{align}
which are identical to those adopted in Ref.~\cite{Fang:2019lmd} and will therefore not be discussed further here.

\subsection{\(U(3)\) extension and determinant interaction}

The scalar field \(X\) is parametrized as
\begin{align}
X = e^{iT^a\eta^a}
\left(\langle X\rangle + S^a T^a\right)
e^{iT^a\eta^a},
\qquad a=0,\dots,8 ,
\end{align}
where \(T^a\) are the generators of \(U(3)\). They are defined as
\begin{align}
T^0=\frac{1}{\sqrt{6}}\operatorname{diag}(1,1,1),
\qquad
T^a=\frac{1}{2}\lambda^a ,
\end{align}
with \(\lambda^a\) denoting the Gell-Mann matrices. With this choice the generators satisfy the normalization condition
\(\operatorname{Tr}(T^a T^b)=\delta^{ab}/2\).
The fields \(\eta^a\) and \(S^a\) represent pseudoscalar and scalar fluctuations, respectively.

The vacuum expectation value (VEV) of \(X\) is taken as
\begin{align}
\langle X\rangle
= \frac{1}{\sqrt{2}}
\operatorname{diag}(\chi_u,\chi_d,\chi_s),
\qquad
\chi_u=\chi_d\neq\chi_s ,
\end{align}
which allows for explicit flavor-symmetry breaking between the light and strange quark sectors.

To model explicit axial \(U(1)_A\) symmetry breaking, we include a determinant interaction of the form
\begin{align}
-\gamma(z)\operatorname{Re}\{\det X\}.
\end{align}
A term of this type was already introduced in Refs.~\cite{Fang:2018vkp,Fang:2019lmd} as a phenomenological ingredient to improve the description of chiral thermodynamics \cite{Chelabi:2015cwn,Chelabi:2015gpc}. In those works, however, the determinant coupling was assumed to be constant, and the analysis focused primarily on the properties of the chiral transition. The dynamical pseudoscalar singlet sector, including \(\eta\)-\(\eta^\prime\) mixing and the generation of the \(\eta^\prime\) mass, was not explicitly addressed.

In the present work, the determinant coupling is generalized to a holographic-coordinate-dependent profile \(\gamma(z)\), allowing for a more flexible realization of the effective \(U(1)_A\)-breaking strength across energy scales. Such a scale- or temperature-dependent realization of the
effective $U(1)_A$-breaking interaction has also been widely
employed in effective QCD models, where the anomaly strength
is often treated as a running or temperature-dependent
parameter \cite{Pisarski:1983ms,Schaffner-Bielich:1999cux,Fukushima:2001hr,Costa:2005cz,Pisarski:2024esv}. Because vacuum spectroscopy alone does not uniquely determine the functional form of \(\gamma(z)\), we consider representative ansätze that provide comparable descriptions of the vacuum pseudoscalar sector.

In particular, we study three classes of profiles,
\begin{align}
\text{Type A:}\quad 
\gamma_A(z) &= \gamma_0 z^2,
\\
\text{Type B:}\quad 
\gamma_B(z) &= \frac{\gamma_0 z^2}{1+\gamma_1 z^2},
\\
\text{Type C:}\quad 
\gamma_C(z) &= \gamma_0 z^2 e^{-\gamma_1 z^2},
\end{align}
where \(\gamma_0\) and \(\gamma_1\) are phenomenological parameters. 
The Type-A profile represents the simplest quadratic realization, in which the anomaly strength increases monotonically toward the infrared. The Type-B profile introduces infrared saturation, providing a smooth interpolation between quadratic ultraviolet growth and a finite asymptotic value at large holographic coordinate \(z\). In contrast, the Type-C profile exhibits a peaked structure, where the anomaly strength grows at intermediate scales but becomes reduced in the deep infrared. All three profiles satisfy \(\gamma(0)=0\), ensuring that the effective anomaly-induced interaction is suppressed at the ultraviolet boundary.

Within the \(U(3)\) flavor framework, the determinant interaction contributes to both the VEV and fluctuation sectors, dynamically affecting the singlet channel as well as singlet-octet mixing in the pseudoscalar and scalar meson sectors. The profile \(\gamma(z)\) should therefore be regarded as an effective phenomenological function encoding explicit \(U(1)_A\) breaking within the bottom-up holographic description. The parameters associated with each ansatz will be constrained using the vacuum meson spectra, and their implications for the chiral phase structure will also be investigated in the context of the Columbia plot.

\section{Pseudoscalar sector and constraints on the anomaly profile}\label{Sec: Pseudoscalar sector}

In this section, we analyze the pseudoscalar sector of the model and use it to constrain the effective anomaly profile \(\gamma(z)\). We first present the pseudoscalar mass spectrum obtained from the coupled equations of motion and examine the singlet-octet structure of the corresponding holographic wavefunctions. We then compute the decay constants and extract the associated mixing parameters. Finally, we compare different functional forms of \(\gamma(z)\) that yield comparable vacuum descriptions and discuss their implications for the subsequent analysis of the chiral phase structure.

The numerical values of the model parameters used in this work are summarized in Table.~\ref{tab: model parameters}, where A, B, and C correspond to the different anomaly-profile ansätze introduced above. These parameter sets are chosen such that each profile provides a consistent description of the vacuum pseudoscalar observables.

\begin{table}[h]
    \centering
    \caption{Model parameters used in this work.}
    \label{tab: model parameters}
    \resizebox{\columnwidth}{!}{
    \begin{tabular}{cccccccc}
    \hline
    \hline
    Type & \(m_u\)(MeV) & \(m_s\) (MeV) & \(\mu_g\)(MeV) & \(\mu_c\)(MeV) & \(\lambda\) & \(\gamma_0\)(GeV\(^2\)) & \(\gamma_1\)(GeV\(^2\)) \\
    \hline
    $A$ & 3.9 & 110 & 480 & 1300 & 130 & -2.2 & --- \\
    $B$ & 3.8 & 105 & 480 & 1200 & 130 & -70 & 5.7 \\
    $C$ & 3.8 & 105 & 460 & 1200 & 75 & -85 & 1.2 \\
    \hline
    \hline
    \end{tabular}
    }
\end{table}

\subsection{Pseudoscalar spectrum and singlet-octet structure}

The pseudoscalar meson spectrum is obtained by solving the coupled fluctuation equations derived from the quadratic expansion of the action around the vacuum background. The explicit form of the coupled equations used in this work reads
\begin{align}
&\partial_z \left( e^{A-\Phi} \, \partial_z \phi_n^a \right)
+ 2 g_5^2 e^{3A-\Phi} (M_A^2)_{ab} \left( \eta_n^b - \phi_n^b \right) = 0, \label{Eq: EOMOfPseudoScalarMeson1}\\
&\partial_z\left(e^{3A-\Phi}\left(M_A^2\right)_{ab}\partial_z\eta^b_n\right)
+m_n^2e^{3A-\Phi}\left(M_A^2\right)_{ab}\left(\eta^b_n-\phi^b_n\right)\notag\\
&+\frac{3}{2\sqrt{2}}e^{5A-\Phi}\gamma(z)\chi_u^2\chi_s\eta^0_n\delta_{0a}=0,
\label{Eq: EOMOfPseudoScalarMeson2}
\end{align}
where \(a,b=0,\dots,8\) denote flavor components. The mass matrix is given by
\begin{equation}
\resizebox{\columnwidth}{!}{$
\begin{aligned}
&M_A^2 =\\
&\begin{pmatrix}
\frac{1}{3} (2\chi_u^2 + \chi_s^2) & 0 & 0 & \frac{\sqrt{2}}{3} (\chi_u^2 - \chi_s^2) \\
0 & \chi_u^2 \, \mathbf{1}_{3 \times 3} & 0 & 0 \\
0 & 0 & \frac{1}{4} (\chi_u + \chi_s)^2 \, \mathbf{1}_{4 \times 4} & 0 \\
\frac{\sqrt{2}}{3} (\chi_u^2 - \chi_s^2) & 0 & 0 & \frac{1}{3} \left( \chi_u^2 + 2 \chi_s^2 \right)
\end{pmatrix}.
\end{aligned}
$}
\end{equation}

\begin{figure}[htbp]
\centering
    \includegraphics[width=0.49\linewidth]{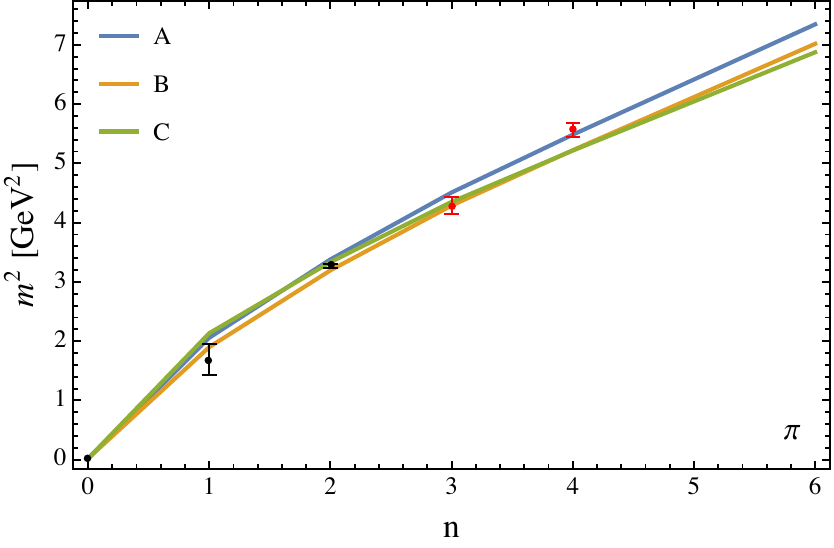}
    \hfill
    \includegraphics[width=0.49\linewidth]{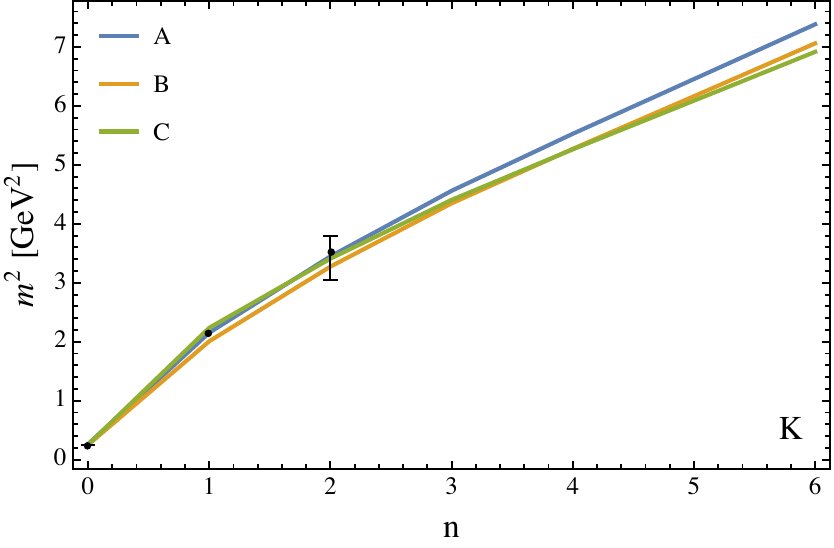}

    \includegraphics[width=0.49\linewidth]{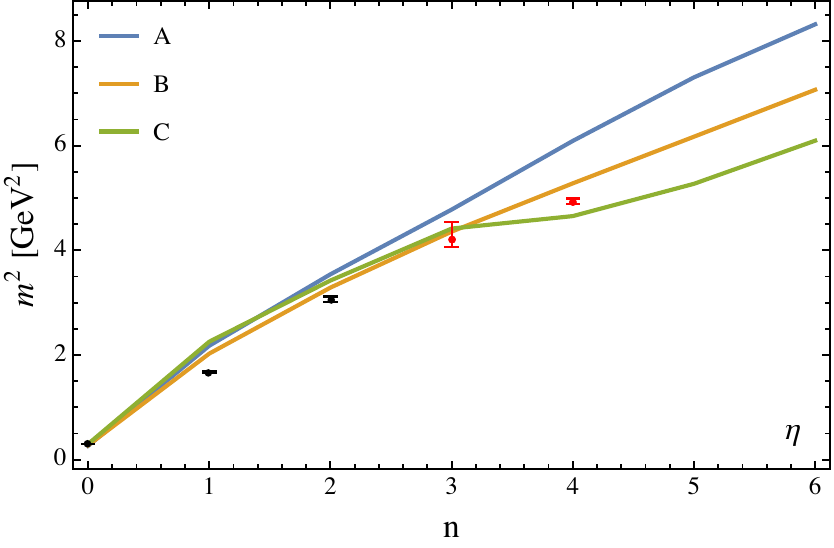}
    \hfill
    \includegraphics[width=0.49\linewidth]{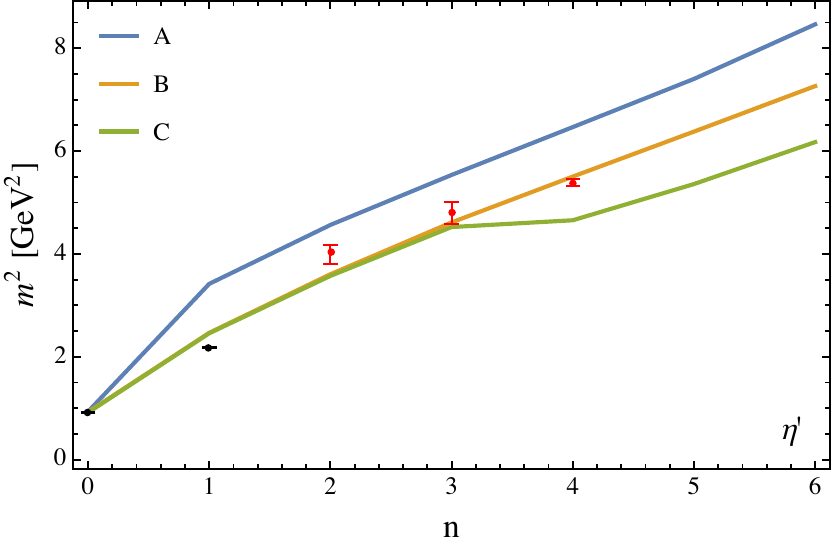}
\caption{Radial trajectories of the pseudoscalar mesons for the three anomaly profiles (A, B, and C). Solid curves denote model predictions, while black points show experimental masses from Ref.~\cite{ParticleDataGroup:2024cfk}. Red points indicate tentative states (marked by \(\dagger\) in Table~\ref{tab: pseudo spectrum}).}
\label{fig: pseudoscalar spectra}
\end{figure}

The pseudoscalar masses are obtained by solving these equations with the boundary conditions
\begin{align}
&\eta_n^a(z \to 0) = \phi_n^a(z \to 0) = 0,\label{Eq: PseudoscalarBC1}\\
&\partial_z \eta_n^a(z \to \infty) = \partial_z \phi_n^a(z \to \infty) = 0.\label{Eq: PseudoscalarBC2}
\end{align}
In this case, the equations of motion for the singlet and octet components (\(\eta^0\) and \(\eta^8\)) are coupled due to the off-diagonal elements of the mass matrix. The resulting pseudoscalar spectrum at the physical point is shown in Table.~\ref{tab: pseudo spectrum} and Fig.~\ref{fig: pseudoscalar spectra}. Some of the higher excited experimental states included for comparison are not yet firmly established. These states are marked by a dagger in the tables and by red points in the figures. They are included only to indicate the possible location of higher excitations and should therefore be interpreted with caution. One can see that the inclusion of the anomaly interaction \(\gamma(z)\) generates the \(\eta\)-\(\eta^\prime\) mass splitting, and the resulting masses agree well with the experimental values.

\begin{table*}[t]
    \centering
    \caption{Pseudoscalar meson spectrum at the physical point for representative anomaly-profile parameter sets. Experimental values from Ref.~\cite{ParticleDataGroup:2024cfk} are shown for comparison. Experimental masses marked with a dagger correspond to states with limited experimental confirmation or uncertain classification.}
    \label{tab: pseudo spectrum}
    \begin{tabular}{ccccccccccccccccc}
\hline
\hline
n & \(\pi\) exp.(MeV) & A & B & C & \(K\) exp.(MeV) & A & B & C \\
\hline
0 & \(139.57\) & 139.6 & 139.5 & 139.2 & \(493.677\) & 514.2 & 500.7 & 505.2 \\
1 & \(1300\pm100\) & 1434.5 & 1378.2 & 1460.8 & \(1460\) & 1467.4 & 1416.9 & 1495.3 \\
2 & \(1810^{+9}_{-11}\) & 1838.3 & 1787.5 & 1826.1 & \(1874^{+73}_{-123}\) & 1856.8 & 1809.5 & 1846.4 \\
3 & \(2070\pm35^\dagger\) & 2123.5 & 2070.9 & 2085.7 & \(\cdots\) & 2135.3 & 2084.9 & 2098.9 \\
4 & \(2360\pm25^\dagger\) & 2342.0 & 2285.3 & 2283.9 & \(\cdots\) & 2350.8 & 2295.8 & 2294.0 \\
5 & \(\cdots\) & 2533.0 & 2473.6 & 2458.7 & \(\cdots\) & 2540.5 & 2482.6 & 2467.3 \\
6 & \(\cdots\) & 2710.5 & 2649.2 & 2622.1 & \(\cdots\) & 2717.1 & 2657.1 & 2629.8 \\
\hline
\hline
n & \(\eta\) exp.(MeV) & A & B & C & \(\eta^\prime\) exp.(MeV) & A & B & C \\
\hline
0 & \(547.862\) & 514.3 & 523.3 & 546.9 & \(957.78\) & 967.1 & 962.9 & 956.7 \\
1 & \(1294\pm4\) & 1475.6 & 1423.7 & 1501.4 & \(1476\pm4\) & 1848.9 & 1570.5 & 1567.8 \\
2 & \(1751\pm15\) & 1882.1 & 1813.3 & 1850.3 & \(2010^{+35\dagger}_{-60}\) & 2136.3 & 1898.8 & 1890.2 \\
3 & \(2050^{+81\dagger}_{-35}\) & 2186.0 & 2087.5 & 2101.5 & \(2190\pm50^\dagger\) & 2352.9 & 2147.8 & 2126.7 \\
4 & \(2221^{+13\dagger}_{-10}\) & 2468.1 & 2297.9 & 2157.4 & \(2320\pm15^\dagger\) & 2542.8 & 2346.0 & 2157.5 \\
5 & \(\cdots\) & 2702.9 & 2484.3 & 2296.0 & \(\cdots\) & 2720.8 & 2525.6 & 2314.9 \\
6 & \(\cdots\) & 2884.3 & 2658.6 & 2469.0 & \(\cdots\) & 2909.3 & 2695.0 & 2484.9 \\
\hline
\hline
\end{tabular}
\end{table*}

\begin{figure}[htbp]
\centering
    \includegraphics[width=0.49\linewidth]{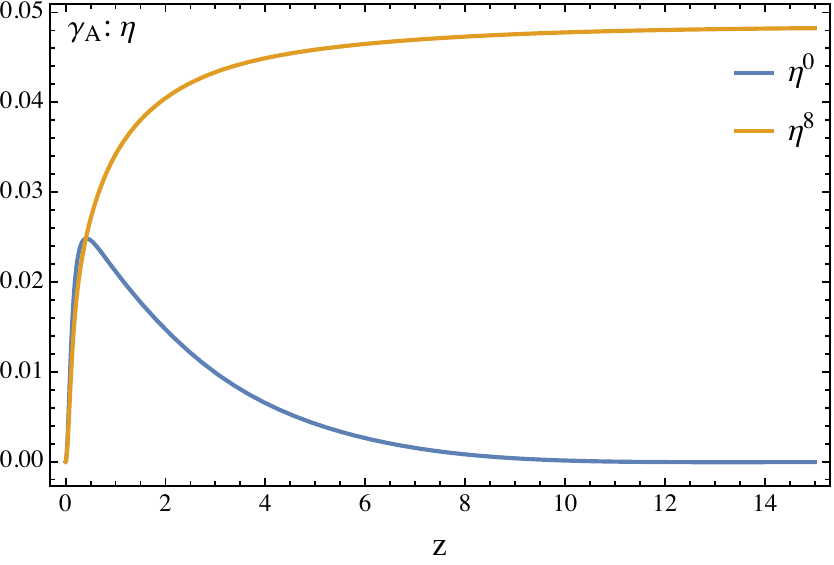}
    \hfill
    \includegraphics[width=0.49\linewidth]{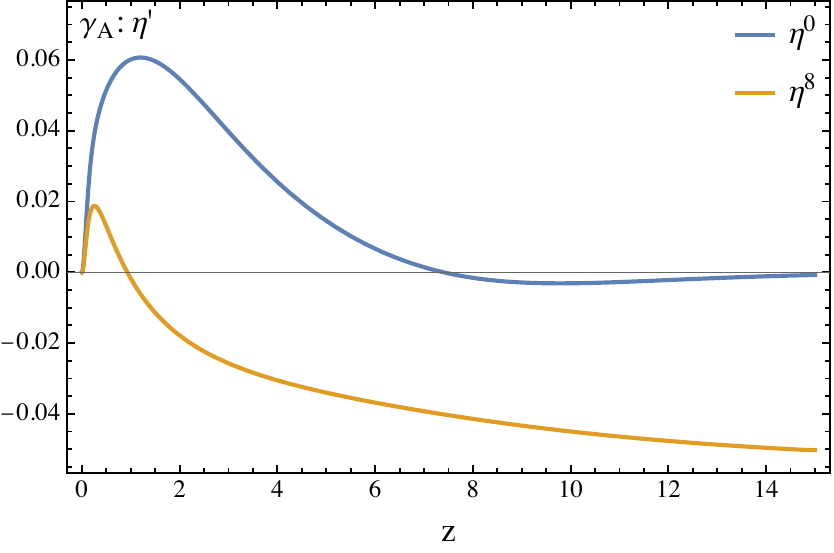}

    \includegraphics[width=0.49\linewidth]{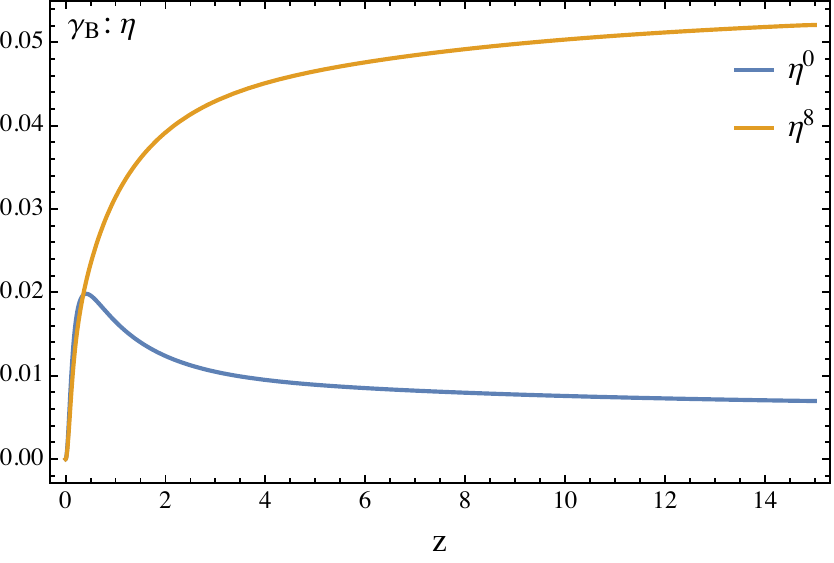}
    \hfill
    \includegraphics[width=0.49\linewidth]{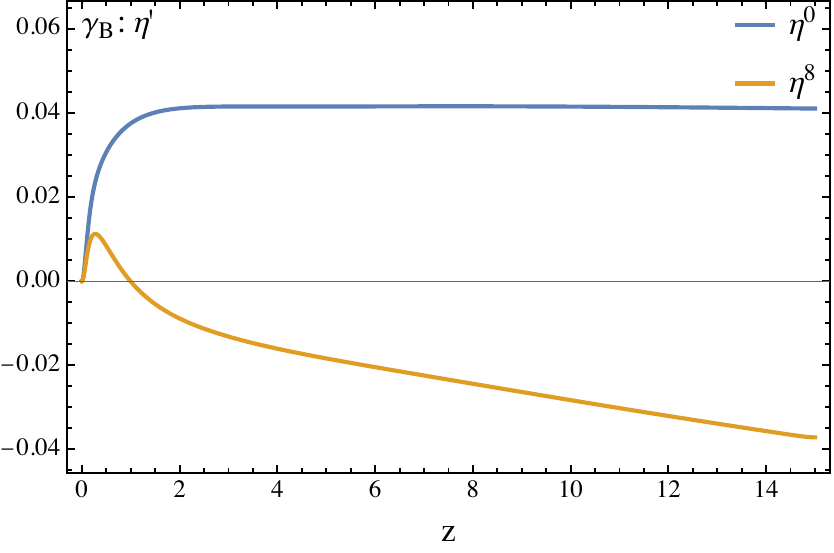}

    \includegraphics[width=0.49\linewidth]{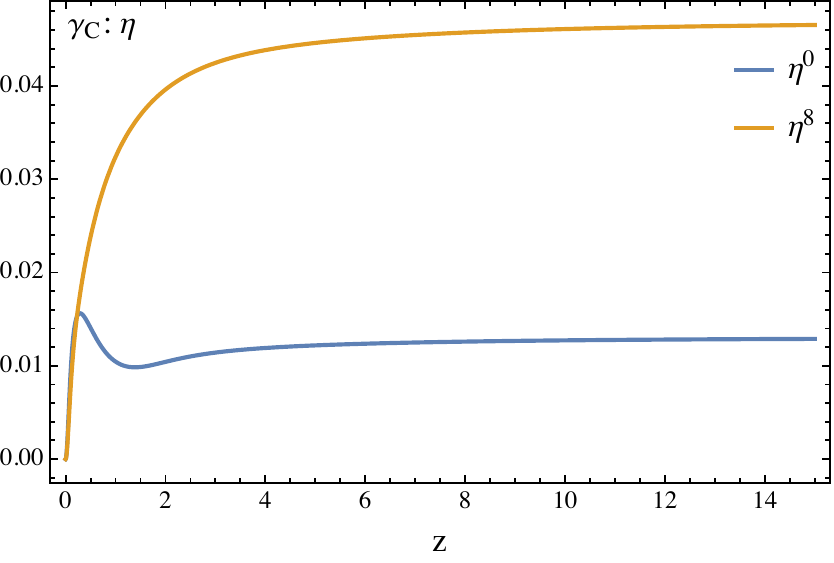}
    \hfill
    \includegraphics[width=0.49\linewidth]{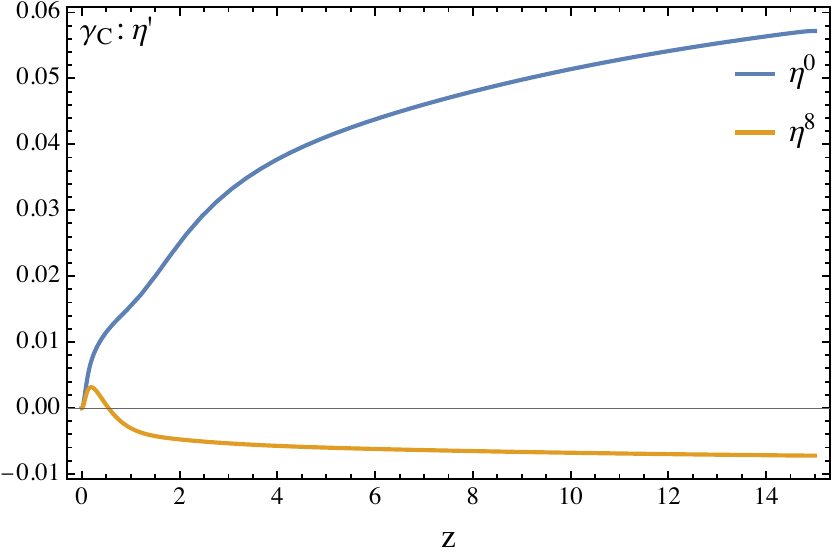}
\caption{Radial wavefunctions of the singlet (\(0\)) and octet (\(8\)) pseudoscalar components for the \(\eta\) and \(\eta^\prime\) states obtained with different anomaly profiles.}
\label{fig: wavefunction}
\end{figure}

The mixing structure of the \(\eta\) and \(\eta^\prime\) mesons can be directly visualized from the holographic wavefunctions. In our numerical solutions, the flavor-singlet component dominates in the \(\eta^\prime\) state, while the octet component dominates in the \(\eta\) state. This behavior is illustrated in Fig.~\ref{fig: wavefunction}. 

\subsection{Decay constants and mixing}

To compute the decay constants and the mixing angles, we first normalize the holographic wavefunctions according to the conditions given in Ref.~\cite{Erlich:2005qh}. 
For \(a = 1,2,3\),
\begin{align}
\int dz\left[\frac{e^{A-\Phi}}{g_5^2}\left(\partial_z\phi^{a}_n\right)^2
+2e^{3A-\Phi}\chi_u^2(\eta^a_n-\phi^a_n)^2\right]=1 .
\end{align}
For \(a = 4,5,6\),
\begin{align}
\int dz\left[\frac{e^{A-\Phi}}{g_5^2}\left(\partial_z\phi^{a}_n\right)^2
+\frac{1}{2}e^{3A-\Phi}(\chi_u+\chi_s)^2(\eta^a_n-\phi^a_n)^2\right]=1 .
\end{align}
For the singlet-octet sector \(a = 0,8\),
\begin{align}
    &\int dz\left\{\frac{e^{A-\Phi}}{g_5^2}\left[\left(\partial_z\phi^{0}_n\right)^2+\left(\partial_z\phi^{8}_n\right)^2\right]+2e^{3A-\Phi}\left[\frac{1}{3}\left(2\chi_u^2\right.\right.\right. \notag\\
    &\left.\left.\left. +\chi_s^2\right)\left(\eta^0_n-\phi^0_n\right)^2+\frac{2\sqrt{2}}{3}\left(\chi_u^2-\chi_s^2\right)\left(\eta^0_n-\phi^0_n\right)\left(\eta^8_n-\phi^8_n\right)\right.\right.\notag\\
    &\left.\left.+\frac{1}{3}\left(\chi_u^2+2\chi_s^2\right)\left(\eta^8_n-\phi^8_n\right)^2\right]\right\} =1.
\end{align}

The decay constants are then computed from
\begin{align}\label{Eq: decay constants}
    f_a = \frac{1}{g_5^2}e^{A-\Phi}\phi_a'|_{z\to 0},
\end{align}
where \(\phi_a\) denotes the normalized wavefunction evaluated at the meson mass \(m=m_i\). For example, to compute \(f_{\eta'}^8\) and \(f_{\eta'}^0\), we set \(m=m_{\eta'}\) and solve Eqs.~\eqref{Eq: EOMOfPseudoScalarMeson1}-\eqref{Eq: EOMOfPseudoScalarMeson2} for \(\phi_8\) and \(\phi_0\). The resulting decay constants are summarized in Table~\ref{tab: decay const}. The results confirm that the \(\eta\) meson is predominantly octet-like, while the \(\eta^\prime\) meson is dominated by the singlet component.

\begin{table}[h]
    \centering
    \caption{Decay constants of the pseudoscalar mesons. Experimental values are taken from Ref.~\cite{ParticleDataGroup:2024cfk}.}
    \label{tab: decay const}
    \begin{tabular}{ccccccccc}
    \hline
    \hline
    Type & \(f_\pi\) & \(f_K\) & \(f_{\eta}^8\) & \(f_{\eta}^0\) & \(f_{\eta^\prime}^8\) & \(f_{\eta^\prime}^0\)\\
    \hline
    Exp. (MeV) & 92.3 & 110 & --- & --- & --- & ---\\
    
     A (MeV) & 84.42 & 93.96 & 88.17 & 20.86 & -41.17 & 87.39 \\
    
     B (MeV) & 76.21 & 85.47 & 82.81 & 13.14 & -31.50 & 82.01 \\
    
     C (MeV) & 96.49 & 104.8 & 103.1 & 14.47 & -29.66 & 96.13 \\
    \hline
    \hline
    \end{tabular}
\end{table}

The mixing angles are extracted from the decay-constant matrix using the two-angle mixing scheme \cite{Feldmann:1998vh,Bickert:2016fgy},
\begin{align}
\begin{pmatrix}
f_{\eta}^8 & f_{\eta}^0 \\
f_{\eta'}^8 & f_{\eta'}^0
\end{pmatrix}
=
\begin{pmatrix}
f_8 \cos \theta_8 & - f_0 \sin \theta_0 \\
f_8 \sin \theta_8 & f_0 \cos \theta_0
\end{pmatrix}.
\end{align}
From this matrix we obtain
\begin{align}
    &f_8=\sqrt{(f_\eta^8)^2+(f_{\eta^\prime}^8)^2},\quad \tan\theta_8=\frac{f_{\eta^\prime}^8}{f_{\eta}^8},\label{Eq: mixing angle1}\\
    &f_0=\sqrt{(f_\eta^0)^2+(f_{\eta^\prime}^0)^2},\quad \tan\theta_0=-\frac{f_{\eta}^0}{f_{\eta^\prime}^0}.\label{Eq: mixing angle2}
\end{align}
The resulting values of \(f_8\), \(f_0\), and the mixing angles are summarized in Table.~\ref{tab: mixing}. For comparison, we also list the best-fit values from phenomenological analyses \cite{Feldmann:1999uf} and results from a lattice QCD study \cite{Bali:2021qem}.
 
\begin{table}[h]
    \centering
    \caption{Values of \(f_8\), \(f_0\), and the mixing angle for different anomaly-profile parameter sets.}
    \label{tab: mixing}
    \begin{tabular}{ccccc}
    \hline
    \hline
    Type & \(f_8\) (MeV) & \(f_0\) (MeV) & \(\theta_8\) & \(\theta_0\) \\
    \hline
    Ref.~\cite{Feldmann:1999uf} & \(116 \pm 4\) & \(108 \pm 3\) & \(-21.2^\circ\pm1.6^\circ\) & \(-9.2^\circ\pm1.7^\circ\) \\
    Ref.~\cite{Bali:2021qem} & \(115 \pm 2.8\) & \(106 \pm 3.2\) & \(-25.8^\circ\pm2.3^\circ\) & \(-8.1^\circ\pm1.8^\circ\) \\
     A & 97.31 & 89.85 & \(-25.03^\circ\) & \(-13.42^\circ\) \\
     B & 88.60 & 83.06 & \(-20.83^\circ\) & \(-9.10^\circ\) \\
     C & 107.28 & 97.21 & \(-16.05^\circ\) & \(-8.56^\circ\) \\
    \hline
    \hline
    \end{tabular}
\end{table}

\subsection{Constraints on the anomaly profile}

We now compare the different anomaly-profile classes introduced in Sec.~\ref{Sec: Introduce the model}. We find that all three functional forms of \(\gamma(z)\) can reproduce the essential features of the vacuum pseudoscalar sector, including the \(\eta\)-\(\eta^\prime\) mass splitting and the singlet-octet mixing pattern, within reasonable parameter ranges. Although quantitative differences appear in specific observables such as decay constants, the overall pseudoscalar spectroscopy can be consistently described across the three classes.

\begin{figure}[htbp]
\centering
    \includegraphics[width=0.7\linewidth]{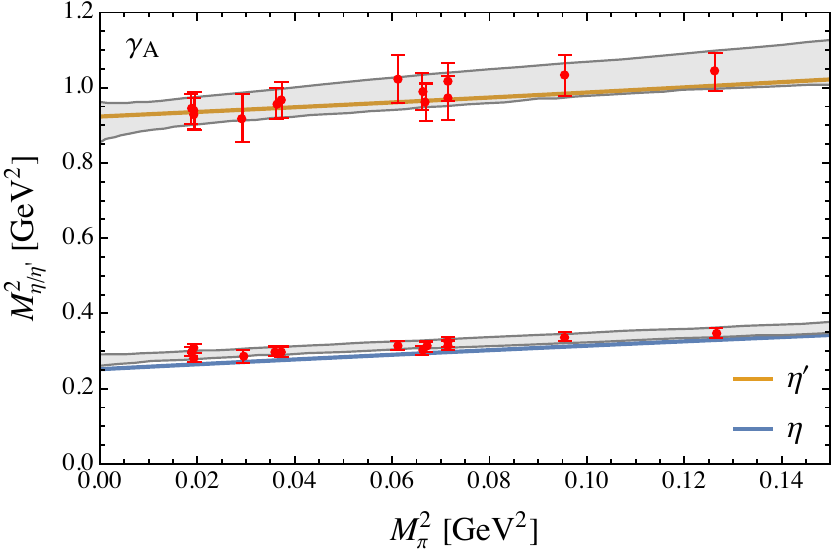}
    \includegraphics[width=0.7\linewidth]{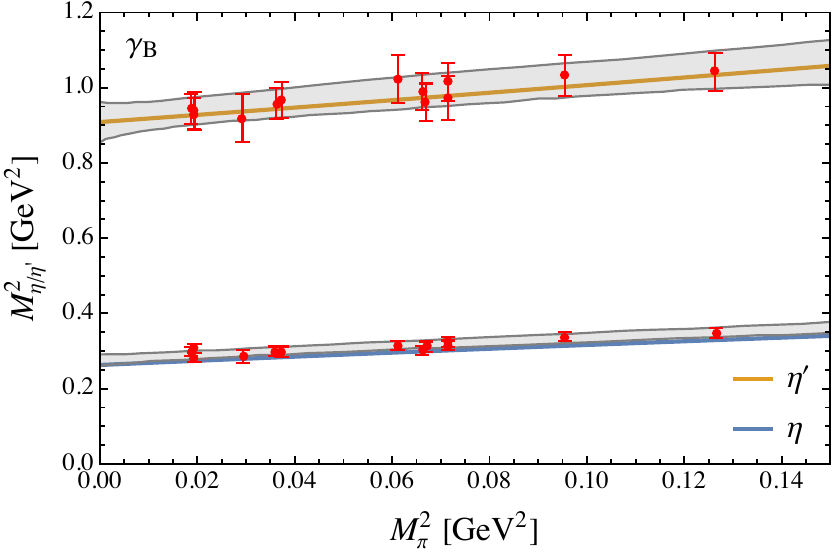}
    \includegraphics[width=0.7\linewidth]{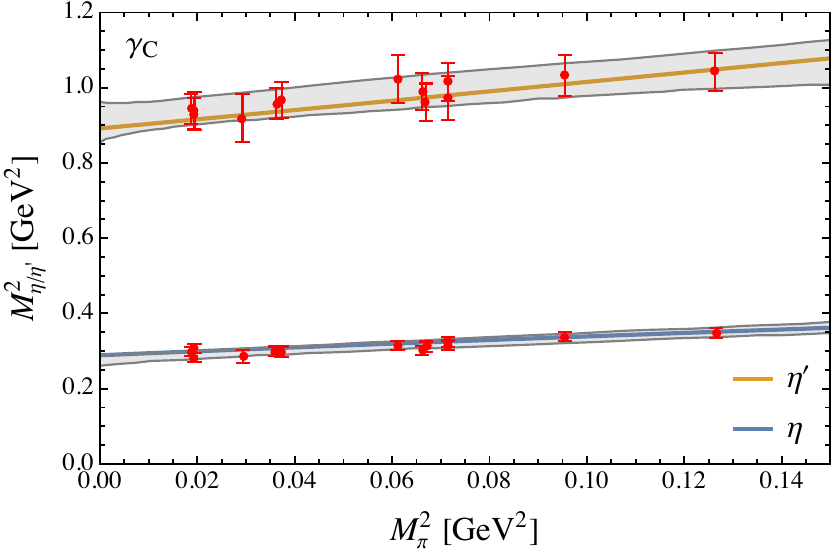}
\caption{Dependence of \(m_{\eta}^2\) and \(m_{\eta'}^2\) on \(m_\pi^2\) along a fixed strange-quark-mass trajectory for representative anomaly-profile parameter sets. Red error bars and gray bands denote lattice-QCD results from Ref.~\cite{Ottnad:2025zxq}.}
\label{fig: mass scan}
\end{figure}

To further probe the pseudoscalar sector, we examine the quark-mass dependence of pseudoscalar observables along a fixed strange-quark-mass trajectory. The resulting behavior of \(m_\eta^2\) and \(m_{\eta^\prime}^2\) as functions of \(m_\pi^2\) is shown in Fig.~\ref{fig: mass scan}, together with recent lattice QCD results \cite{Ottnad:2025zxq}.

We find that the holographic results reproduce the qualitative trends observed in lattice QCD, in particular the evolution of the $\eta$ and $\eta'$ masses with increasing light-quark mass. Across the explored mass range, the model curves lie within the lattice uncertainty bands, indicating that the quark-mass dependence of the pseudoscalar sector is reasonably well captured.

At the same time, different anomaly-profile ans\"atze that provide similar descriptions of the vacuum pseudoscalar spectrum also exhibit comparable quark-mass dependence. This suggests that pseudoscalar observables mainly constrain the overall strength of axial $U(1)_A$ symmetry breaking, while they are not sufficient to uniquely determine the detailed functional form of the anomaly profile $\gamma(z)$.

In summary, the pseudoscalar sector successfully reproduces the essential features of the $\eta$-$\eta'$ system and provides meaningful constraints on the magnitude of the anomaly-induced interaction. However, the vacuum spectroscopy and quark-mass dependence alone do not uniquely determine the detailed functional form of the anomaly profile $\gamma(z)$. This ambiguity motivates us to further examine the role of the anomaly profile in other observables that are sensitive to the dynamics of chiral symmetry breaking. In the following sections, we extend the analysis by incorporating scalar meson observables, which provide additional vacuum constraints on the anomaly profile beyond the pseudoscalar sector. We then proceed to investigate the thermodynamic properties of the model and examine how different realizations of $\gamma(z)$ influence the finite-temperature chiral phase structure and the resulting Columbia plot.

\section{Extended vacuum constraints from scalar and pseudoscalar mesons}\label{Sec: Scalar sector}

To further constrain the anomaly profile $\gamma(z)$, it is necessary to consider additional vacuum observables beyond the pseudoscalar sector. While pseudoscalar mesons are directly sensitive to axial $U(1)_A$-breaking effects, scalar mesons probe complementary aspects of the holographic potential and bulk dynamics. A simultaneous description of both scalar and pseudoscalar sectors therefore provides a more stringent and comprehensive test of the anomaly-profile implementation.

In this section, we perform a combined analysis of scalar and pseudoscalar mesons to investigate whether a single parameter set can reproduce observables in both sectors. This allows us to evaluate how strongly the anomaly profile can be constrained once multiple classes of vacuum observables are incorporated.

\subsection{Simultaneous fit to scalar and pseudoscalar spectra}

The scalar meson spectrum is obtained by solving the linearized equations of motion for scalar fluctuations around the vacuum background,
\begin{align}
    \partial_z\left(e^{3A-\Phi}\partial_zS^a_n\right)-e^{5A-\Phi}\left(M_S^2\right)_{ab}S^b_n+m_n^2e^{3A-\Phi}S^a_n =0,
\end{align}
with
\begin{align}
M_S^2 =\begin{pmatrix}
A & 0 & 0 & B \\
0 & C \mathbf{1}_{3 \times 3} & 0 & 0 \\
0 & 0 & D \mathbf{1}_{4 \times 4} & 0 \\
E & 0 & 0 & F
\end{pmatrix}.
\end{align}
Here the matrix elements are given by
\begin{align}
    &A = m_5^2+\frac{\gamma\left(2\chi_u+\chi_s\right)}{3\sqrt{2}}+\lambda\left(2\chi_u^2+\chi_s^2\right),\\
    & B = \frac{\gamma\left(\chi_s-\chi_u\right)}{6}+\sqrt{2}\lambda\left(\chi_u^2-\chi_s^2\right),\\
    & C = m_5^2-\frac{\gamma\chi_s}{2\sqrt{2}}+3\lambda\chi_u^2,\\
    & D = m_5^2-\frac{\gamma\chi_u}{2\sqrt{2}}+\lambda\left(\chi_u^2+\chi_u\chi_s+\chi_s^2\right),\\
    & E = \frac{\gamma\left(\chi_s-\chi_u\right)}{6}+\sqrt{2}\lambda\left(\chi_u^2-\chi_s^2\right),\\
    & F = m_5^2+\frac{\gamma\left(\chi_s-4\chi_u\right)}{6\sqrt{2}}+\lambda\left(\chi_u^2+2\chi_s^2\right).
\end{align}
The corresponding boundary conditions are
\begin{align}
&S_n^a(z \to 0) = 0,\label{Eq: ScalarBC1}\\
&\partial_z S_n^a(z \to \infty) = 0.\label{Eq: ScalarBC2}
\end{align}

\begin{figure}[htbp]
    \centering
    \includegraphics[width=0.49\linewidth]{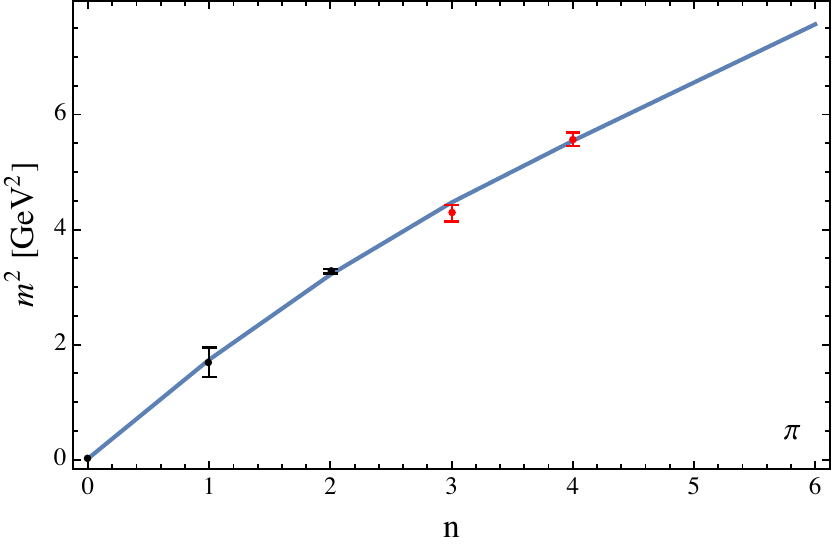}
    \hfill
    \includegraphics[width=0.49\linewidth]{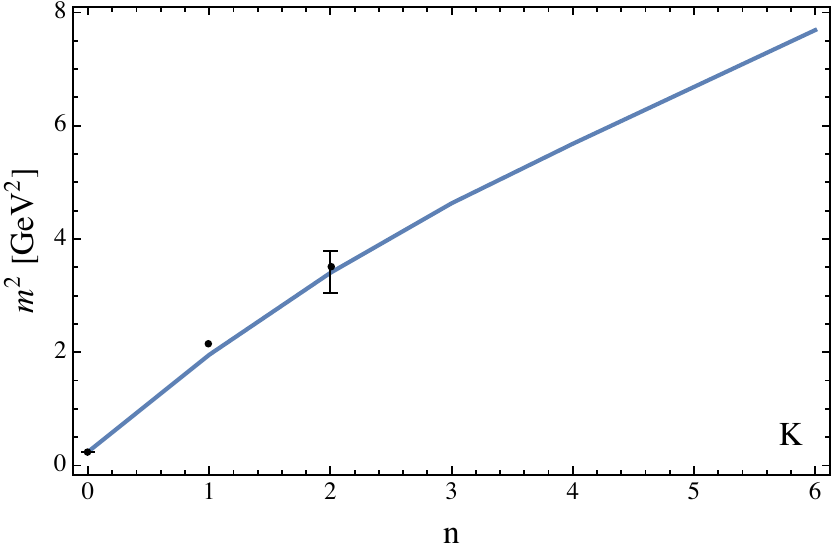}
    \includegraphics[width=0.49\linewidth]{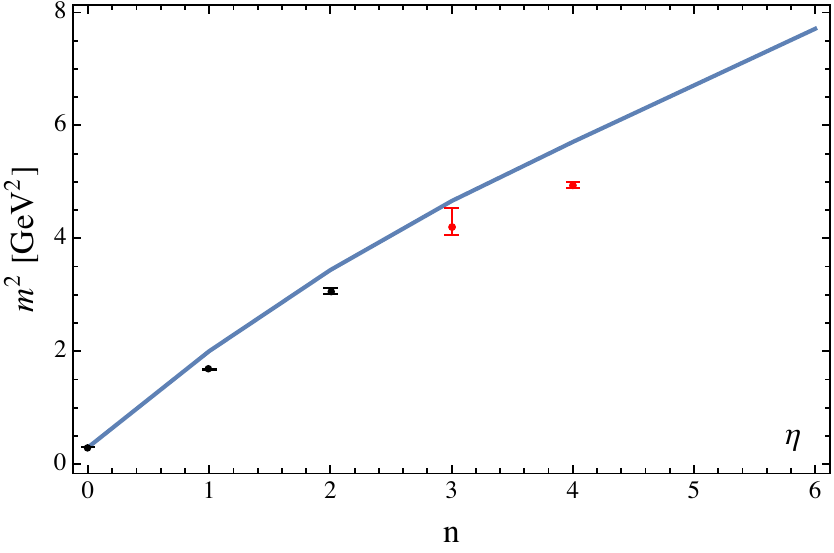}
    \hfill
    \includegraphics[width=0.49\linewidth]{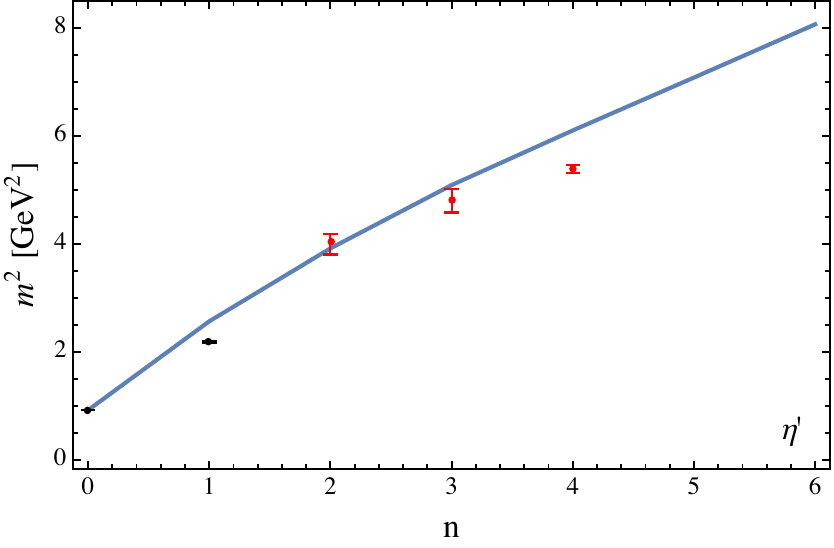}
    \includegraphics[width=0.49\linewidth]{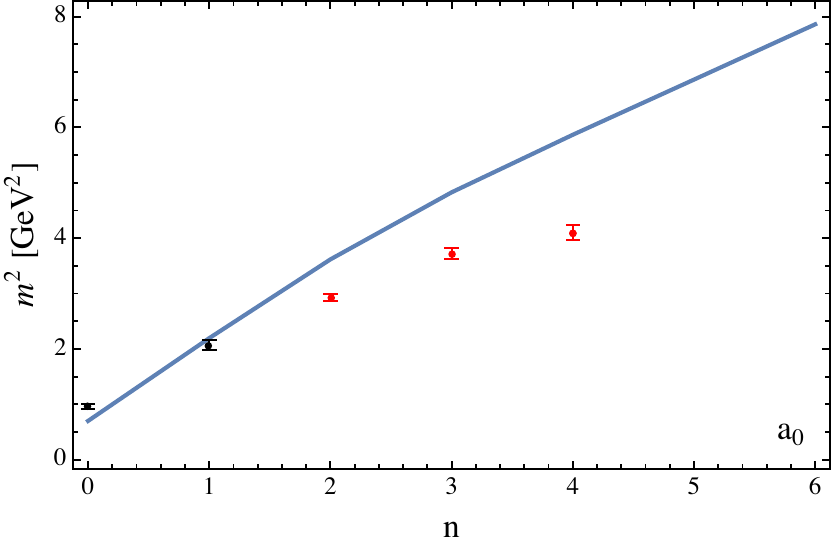}
    \hfill
    \includegraphics[width=0.49\linewidth]{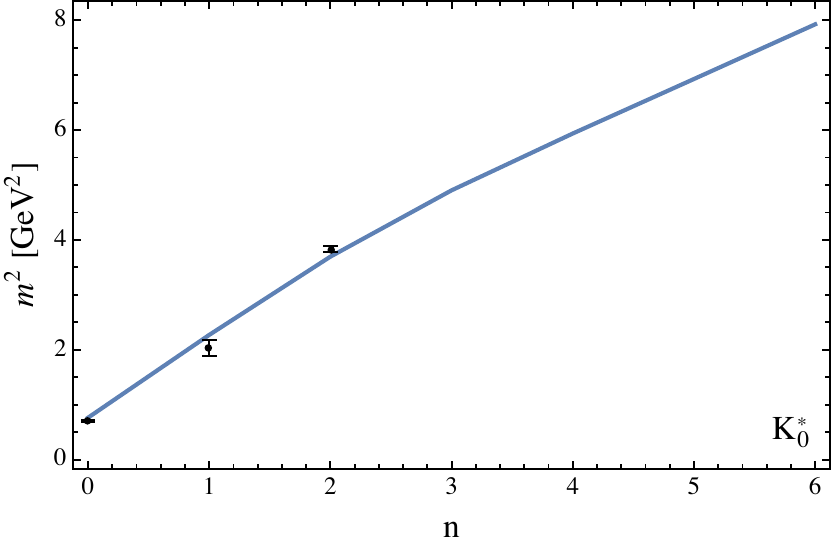}
    \includegraphics[width=0.49\linewidth]{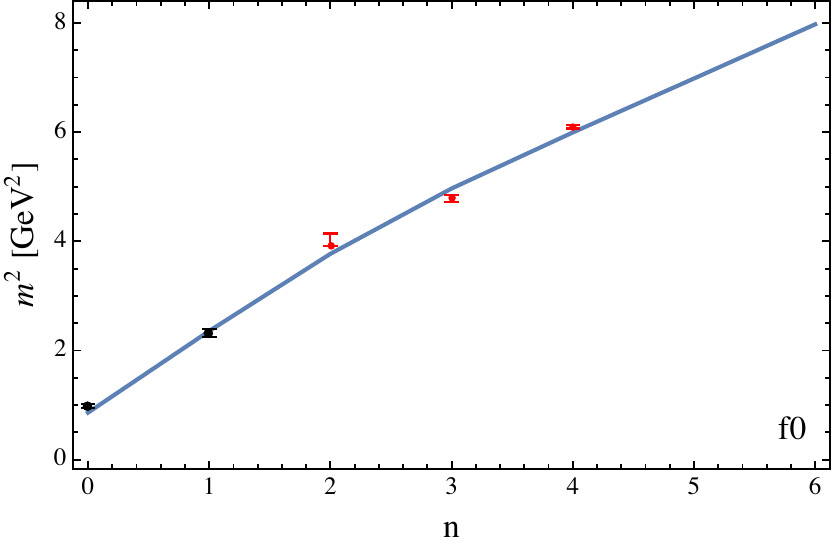}
    \hfill
    \includegraphics[width=0.49\linewidth]{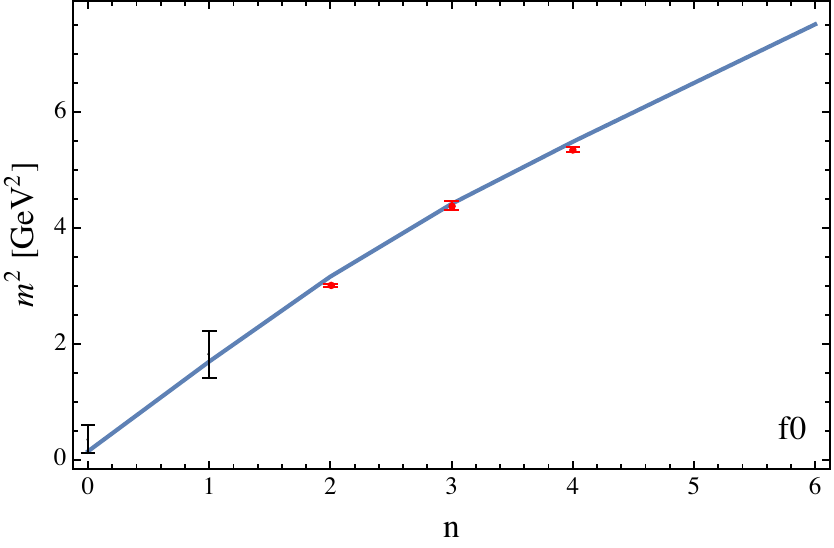}
\caption{
Radial trajectories of pseudoscalar and scalar mesons from the simultaneous scalar-pseudoscalar fit with the Type-B anomaly profile.
Solid curves denote model predictions and points show experimental masses from Ref.~\cite{ParticleDataGroup:2024cfk}.
Red points indicate tentative states (marked by $\dagger$ in Table~\ref{tab: ScalarSpectra}).
}
    \label{fig: pseudoscalar and scalar spectra}
\end{figure}

Meanwhile, the pseudoscalar meson spectrum continues to be determined from Eqs.~\eqref{Eq: EOMOfPseudoScalarMeson1}--\eqref{Eq: EOMOfPseudoScalarMeson2} together with the boundary conditions in Eqs.~\eqref{Eq: PseudoscalarBC1} and \eqref{Eq: PseudoscalarBC2}.
We then perform a simultaneous fit to the scalar and pseudoscalar meson spectra by adjusting the model parameters within each anomaly-profile ansatz. The fitted parameter set is listed in Table~\ref{tab: CombinedFitParameters}. 
\begin{table}[h]
    \centering
    \caption{Model parameters from a simultaneous scalar-pseudoscalar fit using the Type-B anomaly profile.}
    \label{tab: CombinedFitParameters}
    \resizebox{\columnwidth}{!}{
    \begin{tabular}{cccccccc}
    \hline
    \hline
    Type & \(m_u\)(MeV) & \(m_s\) (MeV) & \(\mu_g\)(MeV) & \(\mu_c\)(MeV) & \(\lambda\) & \(\gamma_0\)(GeV\(^2\)) & \(\gamma_1\)(GeV\(^2\)) \\
    \hline
    $B$ & 3.8 & 108 & 504.7 & 900 & 40 & -20 & 0.7 \\
    \hline
    \hline
    \end{tabular}
    }
\end{table}
The resulting scalar and pseudoscalar spectra obtained from the combined fit are summarized in Table~\ref{tab: ScalarSpectra} and Fig.~\ref{fig: pseudoscalar and scalar spectra}.
\begin{table*}[t]
    \centering
    \caption{Scalar and pseudoscalar meson spectra from the simultaneous fit, with experimental values from Ref.~\cite{ParticleDataGroup:2024cfk} shown for comparison. Experimental masses marked with a dagger correspond to states with limited experimental confirmation or uncertain classification.}
    \label{tab: ScalarSpectra}
    \begin{tabular}{ccccccccc}
    \hline
    \hline
    \multicolumn{9}{c}{\textbf{Pseudoscalar mesons}} \\
    \hline
    \hline
    n & \(\pi\) exp.(MeV) &  B & \(K\) exp.(MeV) &  B & \(\eta\) exp.(MeV) &  B & \(\eta'\) exp.(MeV) &  B   \\
    \hline
    
    0 & 139.57 & 139.8 & 493.677 & 490.5 & 547.862 & 545.4 & 957.78 & 957.3 \\
    
    1 & \(1300\pm100\) & 1318.8 & 1460 & 1398.0 & \(1294\pm4\) & 1414.7 & \(1476\pm4\) & 1600.5 \\
    
    2 & \(1810^{+9}_{-11}\) & 1793.2 & \(1874^{+73}_{-123}\) & 1844.5 & \(1751\pm15\) & 1854.3 & \(2010^{+35\dagger}_{-60}\) & 1978.0 \\

    3 & \(2070\pm35^\dagger\) & 2114.6 & \(\cdots\) & 2151.8 & \(2050^{+81\dagger}_{-35}\) & 2158.6 & \(2190\pm50^\dagger\) & 2254.8\\

    4 & \(2360\pm25^\dagger\) & 2354.2 & \(\cdots\) & 2383.5 & \(2221^{+13\dagger}_{-10}\) & 2388.9 & \(2320\pm15^\dagger\) & 2469.0 \\

    5 & \(\cdots\) & 2560.6 & \(\cdots\) & 2585.6 & \(\cdots\) & 2590.1 & \(\cdots\) & 2659.9\\

    6 & \(\cdots\) & 2751.0 & \(\cdots\) & 2772.9 & \(\cdots\) & 2777.0 & \(\cdots\) & 2839.2\\
    \hline
    \hline
    \multicolumn{9}{c}{\textbf{Scalar mesons}} \\
    \hline
    \hline
    n & \(a_0\) exp.(MeV) &  B & \(K_0^*\) exp.(MeV) &  B & \(f_0\) exp.(MeV) &  B & \(f_0\) exp.(MeV) &  B  \\
    \hline
    
    0 & \(980\pm20\) & 836.3 & \(838\pm11\) & 874.4  & \(990\pm20\) & 928.5 & \(400-800\) & 380.9\\
    
    1 & \(1439\pm34\) & 1481.2 & \(1425\pm50\) & 1507.1  & \(1522\pm25\) & 1535.2 & \(1200 - 1500\) & 1301.7\\
    
    2 & \(1713\pm19^\dagger\) & 1901.7 & \(1957\pm14\) & 1922.5  & \(1982^{+54\dagger}_{-3}\) & 1941.8 & \(1733^{+8\dagger}_{-7}\) & 1777.8\\
    
    3 & \(1931\pm26.1^\dagger\) & 2197.1 & \(\cdots\) & 2214.4 & \(2187\pm14^\dagger\) & 2228.7 & \(2095^{+17\dagger}_{-19}\) & 2101.3\\

    4 & \(2025\pm35^\dagger\) & 2422.1 & \(\cdots\) & 2436.7 & \(2470^{+5.7\dagger}_{-7.2}\) & 2448.2 & \(2312^{+10\dagger}_{-7.6}\) & 2342.5\\

    5 & \(\cdots\) & 2619.7 & \(\cdots\) & 2632.5 & \(\cdots\) & 2642.3 & \(\cdots\) & 2550.1 \\

    6 & \(\cdots\) & 2803.7 & \(\cdots\) & 2815.1 &\(\cdots\) & 2823.8 & \(\cdots\) & 2741.4 \\
    \hline
    \hline
    \end{tabular}
\end{table*}

We find that among the anomaly-profile ansätze considered in Sec.~\ref{Sec: Pseudoscalar sector}, only the Type-B profile admits a parameter set capable of simultaneously reproducing both scalar and pseudoscalar meson spectra with reasonable accuracy. By contrast, the Type-A and Type-C profiles, while successful in describing pseudoscalar observables, fail to yield satisfactory scalar spectra within the explored parameter ranges. This demonstrates that scalar mesons impose additional nontrivial constraints on the anomaly profile that are not captured by pseudoscalar observables alone.

\subsection{Decay constants, mixing, and quark-mass dependence}

Using the parameter set obtained from the simultaneous scalar and pseudoscalar fit for the Type-B anomaly profile, we compute the pseudoscalar decay constants and mixing parameters using Eq.~\eqref{Eq: decay constants} and Eqs.~\eqref{Eq: mixing angle1}-\eqref{Eq: mixing angle2}. The resulting decay constants are summarized in Table~\ref{tab: ScalarDecayConstant}.
\begin{table}[h]
    \centering
    \caption{Pseudoscalar decay constants obtained from the simultaneous fit, with experimental values from Ref.~\cite{ParticleDataGroup:2024cfk}.}
    \label{tab: ScalarDecayConstant}
    \begin{tabular}{ccccccc}
    \hline
    \hline
    Type & \(f_\pi\) & \(f_K\) & \(f_{\eta}^8\) & \(f_{\eta}^0\) & \(f_{\eta^\prime}^8\) & \(f_{\eta^\prime}^0\)\\
    \hline
    Exp. (MeV) & 92.3 & 110 & --- & --- & --- & ---\\
    
     B (MeV) & 50.15 & 65.08 & 66.63 & -5.25 & -24.09 & 61.79 \\
    \hline
    \hline
    \end{tabular}
\end{table}
The corresponding singlet-octet mixing parameters are shown in Table~\ref{tab: ScalarMixingAngle}.
\begin{table}[h]
    \centering
    \caption{Singlet-octet mixing parameters obtained from the simultaneous fit.}
    \label{tab: ScalarMixingAngle}
    \begin{tabular}{ccccc}
    \hline
    \hline
    Type & \(f_8\) (MeV) & \(f_0\) (MeV) & \(\theta_8\) & \(\theta_0\) \\
    \hline
    Ref.~\cite{Feldmann:1999uf} & \(116 \pm 4\) & \(108 \pm 3\) & \(-21.2^\circ\pm1.6^\circ\) & \(-9.2^\circ\pm1.7^\circ\) \\
    Ref.~\cite{Bali:2021qem} & \(115 \pm 2.8\) & \(106 \pm 3.2\) & \(-25.8^\circ\pm2.3^\circ\) & \(-8.1^\circ\pm1.8^\circ\) \\
    B & 70.86 & 62.01 & \(-19.88^\circ\) & \(4.85^\circ\) \\
    \hline
    \hline
    \end{tabular}
\end{table}

We find that while the pseudoscalar masses remain in reasonable agreement with experimental values, the corresponding decay constants are significantly smaller than the experimental ones. This indicates that imposing a simultaneous fit to both scalar and pseudoscalar spectra introduces a noticeable tension in the decay-constant sector. A similar pattern was also observed in our previous two-flavor study \cite{Fang:2016nfj}.

\begin{figure}[htbp]
    \centering
    \includegraphics[width=0.49\linewidth]{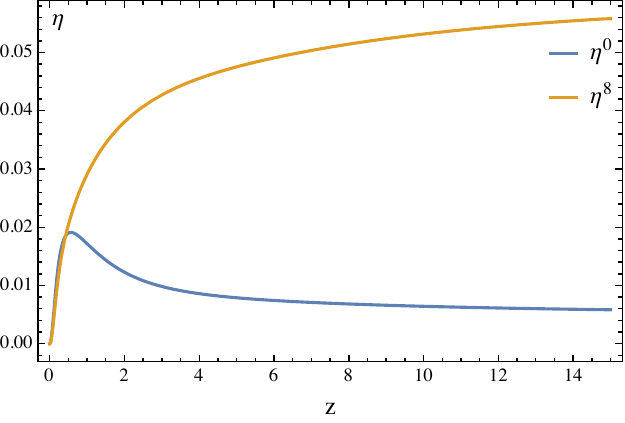}
    \hfill
    \includegraphics[width=0.49\linewidth]{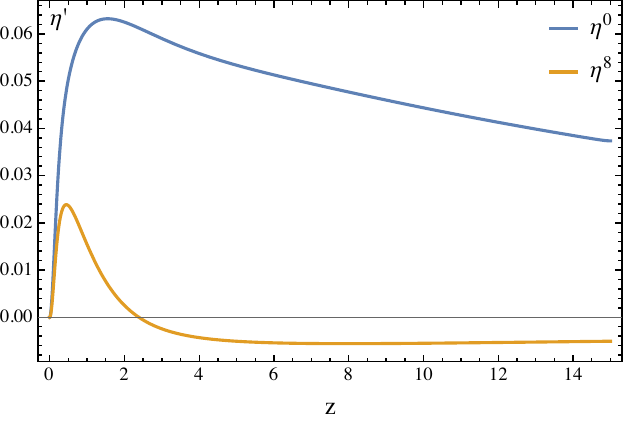}
    \includegraphics[width=0.49\linewidth]{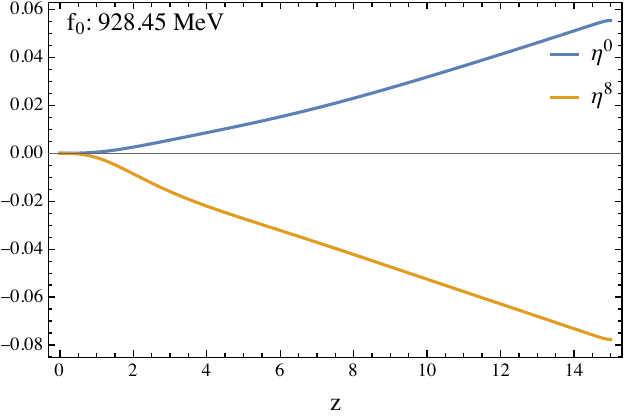}
    \hfill
    \includegraphics[width=0.49\linewidth]{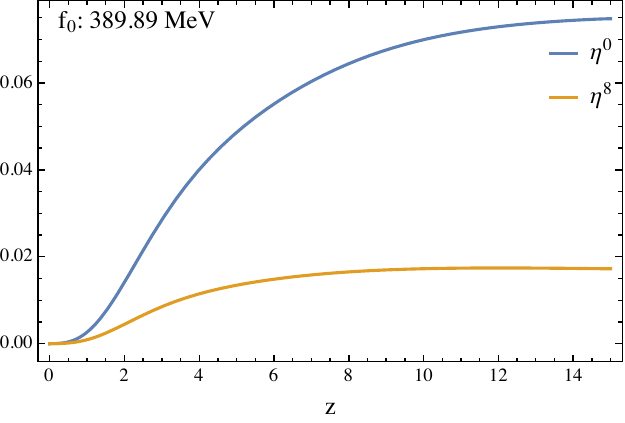}
    \caption{Radial wavefunctions of the singlet (\(a=0\)) and octet (\(a=8\)) components for the two lowest pseudoscalar and scalar eigenstates in the coupled singlet-octet (\(0\)-\(8\)) sector.}
    \label{fig:ScalarWavefunction08}
\end{figure}

Additionally, there is a clear difference between the singlet-octet structures of the pseudoscalar and scalar channels. As shown in Table~\ref{tab: ScalarDecayConstant}, in the pseudoscalar sector the \(\eta\) state is dominated by the \(\eta^8\) component, whereas the \(\eta^\prime\) state is dominated by the \(\eta^0\) component, consistent with the hierarchy expected from axial-anomaly effects \cite{Feldmann:1998sh,Feldmann:1999uf,Amsler:2018zkm}. In contrast, the ordering is reversed in the scalar sector, as illustrated in Fig.~\ref{fig:ScalarWavefunction08}. The figure displays the radial wavefunctions of the singlet (\(a=0\)) and octet (\(a=8\)) components for the two lowest pseudoscalar and scalar eigenstates. As shown in the figure, the lighter scalar state is predominantly singlet, whereas the heavier state is primarily octet, consistent with the expected structure of the scalar sector \cite{Oller:2003vf,Klempt:2021nuf}.

This reversed singlet-octet ordering reflects a nontrivial impact of the $U(1)_A$-breaking determinant interaction on the dynamics of the singlet channel. In particular, it indicates that the anomaly-induced interaction does not merely shift the pseudoscalar singlet mass (leading to the large \(\eta^\prime\) mass), but also modifies the coupled singlet-octet structure in other channels, thereby giving rise to a qualitatively different pattern in the scalar sector.

\begin{figure}[htbp]
    \centering
    \includegraphics[width=0.7\linewidth]{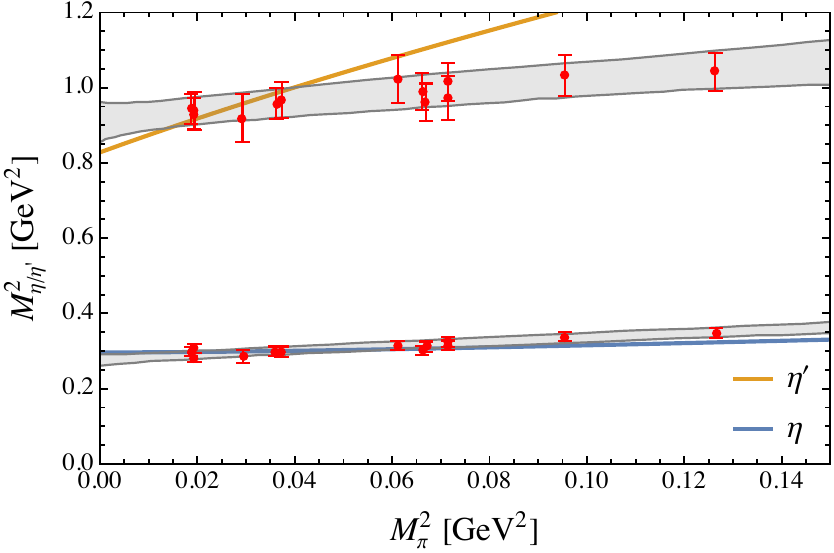}
    \caption{Quark-mass dependence of the \(\eta\) and \(\eta'\) masses obtained from the simultaneous fit.}
    \label{fig:placeholder}
\end{figure}

To further assess the impact of scalar-sector constraints, we examine the quark-mass dependence of pseudoscalar observables. The resulting behavior of \(M_\eta^2\) and \(M_{\eta^\prime}^2\) as functions of \(M_\pi^2\) is shown in Fig.~\ref{fig:placeholder}.
We find that the resulting quark-mass dependence deviates significantly from the trends observed in lattice QCD. In particular, the slope of the predicted line differs from those obtained in Sec.~\ref{Sec: Pseudoscalar sector} with a pseudoscalar-only calibration, indicating that imposing scalar-sector constraints modifies the global vacuum fit in a nontrivial way. While incorporating scalar mesons significantly reduces the allowed parameter space and favors the Type-B anomaly profile, parameter sets that improve the scalar spectra can simultaneously degrade the pseudoscalar decay constants and the quark-mass dependence, reflecting a nontrivial interplay among different observables.





\section{Chiral phase structure and anomaly-profile dependence}\label{Sec: Finite T}

In this section, we investigate the finite-temperature chiral phase structure of the model and analyze its dependence on the anomaly profile \(\gamma(z)\). To this end, we construct the Columbia plot by scanning the light and strange quark masses and determining the nature of the chiral transition from the behavior of the chiral condensates.

\subsection{Finite-temperature setup and chiral observables}

At finite temperature, the background geometry is replaced by the AdS black-hole metric,
\begin{align}
ds^2 = e^{2A(z)} \left( f(z)\,dt^2 - d\vec{x}^{\,2} - \frac{dz^2}{f(z)} \right),
\end{align}
with \(f(z)=1-\frac{z^4}{z_h^4}\). The temperature is determined by
\begin{align}
T = \frac{1}{4\pi} \left| \frac{df}{dz} \right|_{z_h} = \frac{1}{\pi z_h}.
\end{align}

The background scalar fields \(\chi_{u,s}\) are obtained by solving the finite-temperature equations of motion,
\begin{align}
&\chi_u'' 
+ \left( \frac{f'}{f} + 3A' - \Phi' \right) \chi_u'\notag\\
&- \frac{e^{2A}}{f}
\left( m_5^2 \chi_u + \lambda \chi_u^3 + \frac{\gamma}{2\sqrt{2}} \chi_u \chi_s \right) = 0, \\
&\chi_s'' + \left( \frac{f'}{f} + 3A' - \Phi' \right) \chi_s'\notag\\
&- \frac{e^{2A}}{f}\left( m_5^2 \chi_s + \lambda \chi_s^3 + \frac{\gamma}{2\sqrt{2}} \chi_u^2 \right) = 0.
\end{align}

The chiral condensates are extracted from the ultraviolet expansion of the background scalar fields \cite{Fang:2019lmd},
\begin{align}
    \chi_u(z)=&\frac{1}{\sqrt{2}}\left[m_u\zeta z+\frac{\sigma_u}{\zeta}z^3\right.\notag\\
    &\left.+\frac{1}{4}\left(m_u^3\zeta^3\lambda-2m_u\zeta\mu_c^2\right)z^3\log z+\cdots\right],\\
    \chi_s(z)=&\frac{1}{\sqrt{2}}\left[m_s\zeta z+\frac{\sigma_s}{\zeta}z^3\right.\notag\\
    &\left.+\frac{1}{4}\left(m_s^3\zeta^3\lambda-2m_s\zeta\mu_c^2\right)z^3\log z+\cdots\right],\
\end{align}
where \(\sigma\) is identified with the chiral condensate. The order of the chiral transition is determined from the temperature dependence of the condensate: a discontinuity or multi-valued behavior signals a first-order transition, whereas a smooth crossover is characterized by a continuous monotonic variation of the condensate.

\subsection{Chiral transition and Columbia plots}

\begin{figure}[htbp]
\centering
    \includegraphics[width=0.49\linewidth]{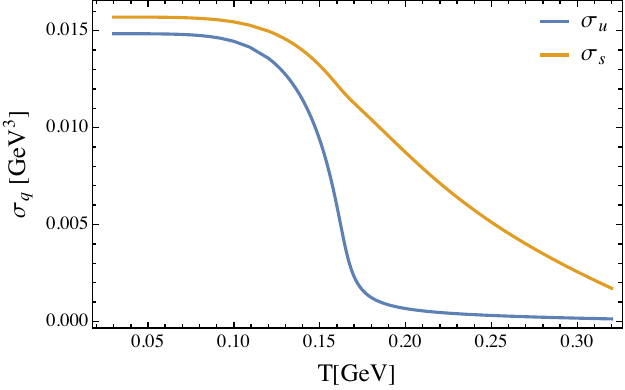}
    \hfill
    \includegraphics[width=0.49\linewidth]{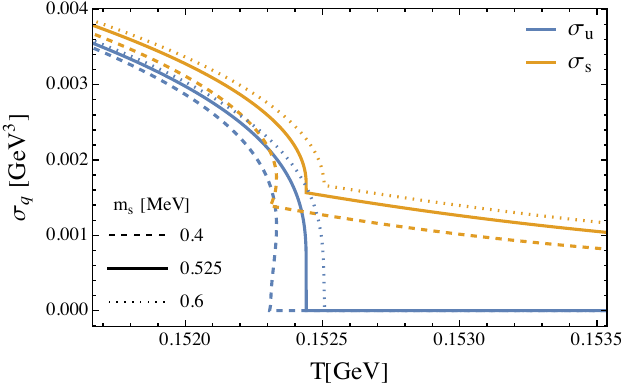}

    \includegraphics[width=0.49\linewidth]{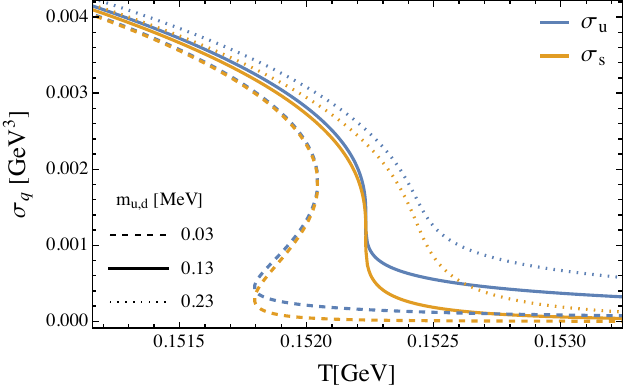}
    \hfill
    \includegraphics[width=0.49\linewidth]{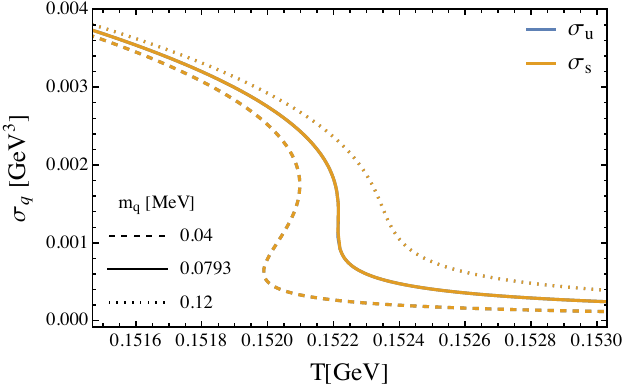}
\caption{Temperature dependence of the quark chiral condensate for the Type-B anomaly profile. Smooth curves indicate crossover behavior, whereas multi-valued branches signal first-order phase transitions. From upper left to lower right: physical quark masses, \(m_{u,d}=0\), \(m_s=0\), and the flavor-symmetric case \(m_{u,d}=m_s\).}
\label{fig: condensate-T}
\end{figure}

To illustrate the nature of the chiral transition for different anomaly profiles, we first examine the temperature dependence of the chiral condensates at several representative quark-mass points. The resulting behavior of the condensates as functions of temperature is shown in Fig.~\ref{fig: condensate-T} for selected parameter sets. For clarity, we focus on the Type-B anomaly profile with parameters fixed from the pseudoscalar-sector fit (Table~\ref{tab: model parameters}) as a representative example.

From these figures we observe that, for the Type-B anomaly profile, the condensate exhibits crossover behavior at the physical quark masses. In the light-quark mass corner, however, a weak multi-branch structure develops, signaling the emergence of a first-order phase transition in that region. Since this feature is not very pronounced, an enlarged view is provided for clarity.

For the Type-A anomaly profile, the chiral condensates vary smoothly with temperature for all explored quark-mass values, with no discontinuities or multi-valued solutions. This behavior indicates a crossover/second-order transition throughout the phase diagram. In contrast, for the Type-C anomaly profile, the qualitative structure of the phase diagram resembles that of Type-B, but with a noticeably larger first-order region.

We also analyze the phase transition behavior for the parameter set introduced in Sec.~\ref{Sec: Scalar sector}. The resulting phase diagram exhibits features qualitatively similar to those obtained for the Type-B and Type-C profiles, but with the largest first-order region among the cases considered.

\begin{figure}[htbp]
\centering
    \includegraphics[width=0.7\linewidth]{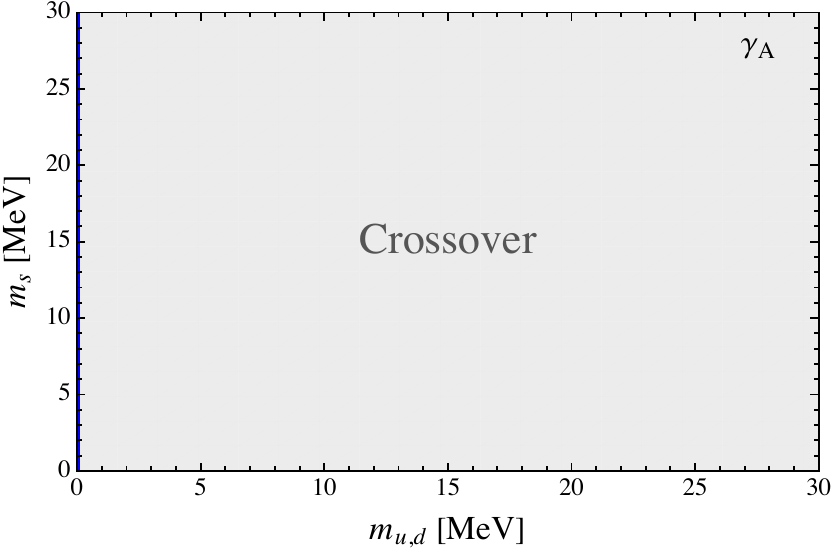}
    
    \includegraphics[width=0.7\linewidth]{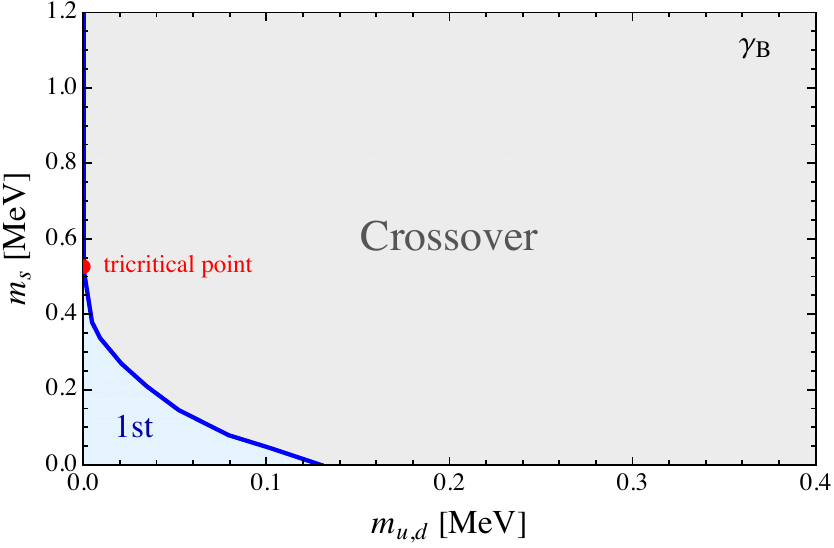}

    \includegraphics[width=0.7\linewidth]{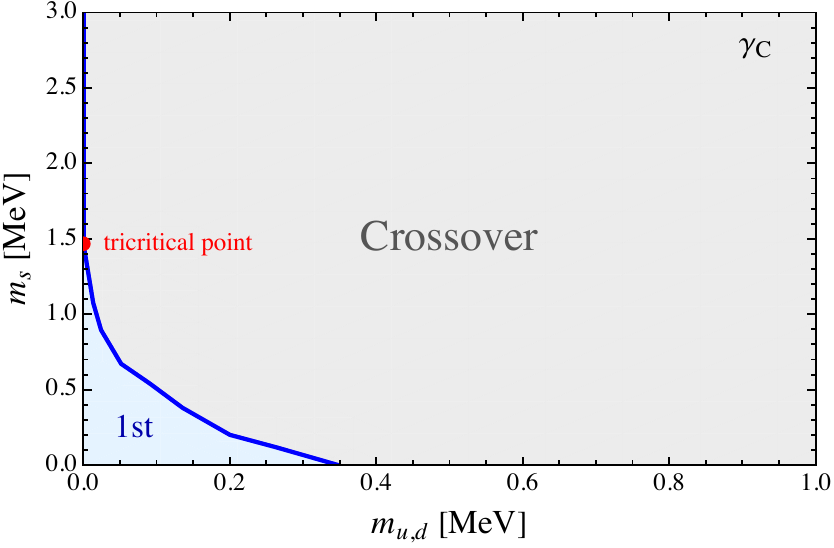}
    
    \includegraphics[width=0.7\linewidth]{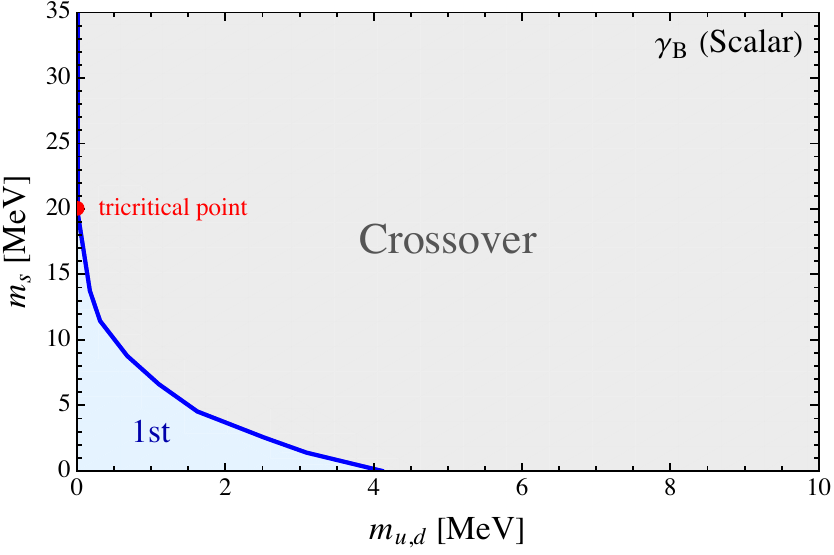}
\caption{Columbia plots obtained for representative anomaly-profile parameter sets. The blue curve denotes the second-order transition line, while the gray and light-blue regions indicate the crossover and first-order transition regions, respectively.}
\label{fig: Columbia}
\end{figure}

Based on this analysis, we construct the Columbia plots by scanning the light and strange quark masses and classifying the transition according to the behavior of the condensates. The resulting Columbia plots are shown in Fig.~\ref{fig: Columbia}. These results demonstrate that the global structure of the Columbia plot is highly sensitive to the specific realization of the anomaly profile, even when different anomaly-profile ansätze provide similarly good descriptions of vacuum pseudoscalar observables.

\subsection{Lattice comparison and implications}

Recent lattice studies have explored the possibility that the first-order region in the Columbia plot may be significantly reduced or even absent in the continuum limit, with the apparent first-order behavior observed at coarse lattice spacing potentially arising from discretization effects \cite{deForcrand:2017cgb,Cuteri:2021ikv,Dini:2021hug}. Motivated by these developments, several effective-field-theory analyses have shown that the predicted chiral phase structure can depend sensitively on how the \(U(1)_A\)-breaking interaction is implemented. In particular, these studies admit a crossover/second-order-type Columbia-plot scenario when anomaly effects are effectively suppressed in the critical region around the chiral transition temperature \cite{Fejos:2022mso,Giacosa:2024orp,Pisarski:2024esv}.

Within our effective \(U(1)_A\) framework, only the Type-A anomaly profile leads to such a crossover/second-order phase structure in the Columbia plot. A possible interpretation is that, for Type-A, the effective anomaly strength becomes significantly weaker near the critical temperature than at low temperature, as reflected by the small ratio \(\gamma_A(z_h(T_c))/\gamma_A(z_h(T_{\mathrm{low}}))\). This behavior suggests that the \(U(1)_A\)-breaking effects are effectively suppressed in the critical region, similar to scenarios explored in some EFT analyses discussed above. By contrast, for the Type-B and Type-C profiles the effective anomaly strength remains comparatively larger in the vicinity of the transition temperature. Within our model, these profiles give rise to a first-order region in the light-quark corner, qualitatively resembling the conventional Columbia-plot scenario \cite{Pisarski:1983ms}.


We emphasize that the present holographic model is not intended to provide a precision quantitative fit to lattice-QCD results. Rather, the comparison should be understood as a qualitative exploration of possible chiral phase structures within an effective strongly coupled framework. 
With this caveat in mind, our results suggest that the anomaly profile can significantly affect the overall structure of the Columbia plot. While vacuum spectroscopy constrains the overall magnitude of \(U(1)_A\)-breaking effects, it does not uniquely determine their detailed implementation. Additional input from finite-temperature and topological observables \cite{Katz:2007tf,Giannuzzi:2021euy} may therefore be required to further constrain the anomaly sector in effective QCD models.

\section{Conclusion and outlook}\label{Sec: Conclusion}

In this work, we constructed a \(U(3)\) extension of a soft-wall holographic QCD model and investigated how different anomaly profiles influence vacuum meson observables and the finite-temperature chiral phase structure. The inclusion of the singlet sector enables a dynamical description of the \(\eta\)-\(\eta^\prime\) system and provides a natural framework for implementing axial \(U(1)_A\)-breaking effects through a holographic-coordinate-dependent determinant interaction.

In the vacuum sector, the model yields a qualitatively consistent description of pseudoscalar masses, decay constants, and singlet-octet mixing. By comparing different functional forms of the anomaly profile \(\gamma(z)\), we found that vacuum pseudoscalar observables constrain the overall magnitude of the anomaly term but do not uniquely determine its detailed form. To further probe the vacuum structure, we also examined the scalar sector for a representative anomaly profile (Type-B). This provides a complementary test of the vacuum fit and shows that, although a reasonable simultaneous description of scalar and pseudoscalar spectra can be achieved in this case, noticeable tensions remain in other observables, particularly the pseudoscalar decay constants and the quark-mass dependence. 

At finite temperature, we constructed the Columbia plot for different anomaly-profile ans\"atze and found that qualitatively different phase structures can emerge from anomaly profiles that provide comparable descriptions of the vacuum sector. In particular, the Type-A profile leads to a crossover/second-order-type pattern, while the Type-B and Type-C profiles produce a first-order region in the light-quark mass corner, consistent with the conventional expectation of a chiral first-order corner. This demonstrates that, within the present holographic framework, the predicted Columbia plot is highly sensitive to the phenomenological implementation of \(U(1)_A\)-breaking effects, and that vacuum constraints alone are insufficient to uniquely determine the resulting global phase structure.

This qualitative picture is broadly consistent with both the conventional Columbia-plot scenario \cite{Pisarski:1983ms} and recent lattice and effective-field-theory studies indicating that the size of the first-order region may depend sensitively on the persistence of axial-anomaly effects near the chiral transition \cite{Fejos:2022mso,Giacosa:2024orp,Pisarski:2024esv}. We emphasize, however, that the present holographic model is not intended to provide a precision quantitative fit to QCD or lattice-QCD data, but rather to serve as a qualitative strongly coupled framework for exploring how different anomaly implementations affect the chiral phase structure.

Future progress will require additional constraints from observables that are more directly sensitive to axial-anomaly effects, such as the topological susceptibility, as well as other finite-temperature probes. It would also be interesting to extend the present analysis to finite baryon chemical potential, where different anomaly profiles may lead to qualitatively different predictions for the phase structure. More generally, clarifying how anomaly-related interactions should be implemented in effective models remains an important step toward improving the reliability of theoretical predictions for the Columbia plot.

\section*{Acknowledgements}
This work is supported by Hunan Provincial Natural Science Foundation of China (Grants No. 2023JJ30115 and No. 2024JJ3004), and also supported by the National Key Research and Development Program of China under Grant No. 2020YFC2201501 and the National Science Foundation of China (NSFC) under Grants No.~12347103 and No.~11821505.

\bibliography{Refs}

\end{document}